\documentstyle[12pt,epsfig]{article} 

\def\frac#1#2{ {{#1} \over {#2} }}
\def\beq{\begin{equation}}
\def\eeq{\end{equation}}
\def\half{\mbox{\small $\frac{1}{2}$}}
\def\O{{\cal O}}

\newcommand{\Tr}[1]{\, {\rm Tr} \, (#1)}
\def\bold#1{\setbox0=\hbox{$#1$}
\kern-.025em\copy0\kern-\wd0
\kern.05em\copy0\kern-\wd0
\kern-.025em\raise.0433em\box0 }
\newcommand{\Fmu}[1]{F_\mu^{(#1)}}
\newcommand{\Amu}[1]{A_\mu^{(#1)}}
\newcommand{\Anu}[2]{A_#1^{(#2)}}
\newcommand{\eq}[1]{Eq.~(\ref{#1})}

\begin{document}

\titlepage 
\begin{flushright}
{\sc uprf-397-1994}
\end{flushright}
\vskip 1.5in
\begin{center}
{\bf \Large Four loop result in $SU(3)$ Lattice Gauge Theory}\\[.1 em]
{\bf \Large by a Stochastic method: 
Lattice correction to the condensate\footnote{
Research partially supported by {\sc murst}, Italy.}
}\\[1.5em]

{\large\bf F.\ Di Renzo, E.\ Onofri}\\ 
\small Dipartimento di Fisica, Universit\`a di Parma and I.N.F.N., Gruppo 
Coll. di Parma\\[.5 em]
{\large\bf  G.\ Marchesini}\\
\small Dipartimento di Fisica, Universit\`a di Milano and 
I.N.F.N., Sezione di Milano\\[.5 em]
\small and\\[.5 em]
{\large\bf  P.\ Marenzoni}\\
\small Dipartimento di Scienze dell'Informazione, Facolt\`a di Ingegneria\\
\small Universit\`a di Parma, 43100 Parma (Italy)\\[1.5 em]
{\small May 1994}\\[2. em]

Abstract
\end{center}
\begin{quote}
We describe a stochastic technique which allows one to compute
numerically the coefficients of the weak coupling 
perturbative expansion of any observable in Lattice  Gauge Theory.
The idea is to insert the exponential representation
of the link variables $U_\mu(x) \to \exp\{A_\mu(x)/\sqrt\beta\}$ 
into the Langevin algorithm and  the 
observables and  to perform the expansion in $\beta^{-\half}$.
The Langevin algorithm is converted into an infinite hierarchy of 
maps which can be exactly truncated at any order.
We give the result for the simple plaquette of $SU(3)$ 
up to fourth loop order ($\beta^{-4}$) which extends by
one loop the previously known series. 
\end{quote}

\newpage

\section{Introduction}
Weak coupling perturbation theory has been developed since
the early times of Lattice Gauge Theory. While the lattice formulation
provides the only method which allows for 
systematic non--perturbative 
calculations from first principles in strongly interacting 
asymptotically free theories, its perturbative expansion is also
important. In particular weak coupling perturbation theory
on the lattice  is useful
in order to accelerate the approach to the continuum, 
like in Symanzik's {\sl improving} program, 
or in the calculation of renormalization constants.
For  observables such as the gluon condensate or the
topological susceptibility 
the weak coupling perturbative expansion on the 
lattice gives an important contribution
to the Monte Carlo data
(see Ref.\cite{DIGIACOMO,BANKS,CPT,CDG}).
The expansion coefficients of the lattice 
correction to the gluon condensate  and the
topological susceptibility have been computed to three loop  order
\cite{ACFP} 
by means of a rather sophisticated diagrammatic technique.

We recently proposed \cite{DALLAS} an alternative technique 
to compute numerically the coefficients of
the expansion of any lattice observable based on 
Parisi--Wu \cite{PARWU}
stochastic quantization implemented on the lattice
\cite{BATROUNI}. The idea is simple: we insert the 
exponential map $U_\mu(x) = \exp\{A_\mu(x)/\sqrt\beta\}$ everywhere in
the elementary Langevin evolution step.
The stochastic evolution of the Lie--algebraic field $A_\mu$
depends parametrically on $\beta$. We expand in powers
of $1/\sqrt\beta$ and obtain the evolution in the Langevin time
order by order. By expanding the observable under consideration,  the
plaquette variable in the present study,
it turns out that the coefficients of 
the perturbative expansion are given by expectation values
of composite operators and are estimated as an  average on the
Langevin history. 

A fundamental aspect of the approach
is concerned with the problem of gauge fixing. We have found
that the Langevin evolution with no gauge fixing is
affected by a random fluctuation increasing in time
and that this problem is taken care of by adopting
a {\sl stochastic gauge fixing} idea proposed by Zwanziger
\cite{ZWANZIGER}
and already implemented on the lattice \cite{ROSSI}.

In this paper we present an application of the method 
by computing the four loop coefficient of the plaquette 
expectation value which is related to the expectation value
of the condensate ${\alpha\over\pi} \langle\bold{F}^2(x)\rangle$. The
first three coefficients agree with the known values.

We present the algorithm in some detail in section two.
In section three we recall the stochastic gauge fixing
method which is needed to keep statistical fluctuations
finite. In section four we report 
the results obtained for the simple plaquette up to fourth 
loop order. In the final section we discuss the possibility 
of extracting the condensate from the plaquette Monte Carlo data 
and we comment on a possible scenario of high order
perturbation theory.

\newpage
\section{The algorithm}\label{sec:algo}
We consider the standard pure $SU(N)$ gauge lattice action 
\begin{equation}
S[U] = -{\beta\over 2N}\sum_P  \Tr{U_P+U_P^\dagger}
\end{equation}
where the sum extends to all plaquettes $P$ in a hypercubic
lattice. We shall denote by $U_\mu(x)$ the link variable at
the site $x$ in the direction $\mu$.
We borrow the implementation of the Langevin dynamics from
Ref.\cite{BATROUNI}. A single Langevin step is given by 
a sweep of the lattice where each link variable is updated
according to the rule
\begin{equation}\label{lange}
U_\mu(x) \to U_\mu'(x) = e^{-F_\mu(x)} \; U_\mu(x)
\end{equation}
where $F_\mu(x) $ is given by
\begin{equation}
F_\mu(x) ={\epsilon\beta\over 4 N} \sum_{P\supset \mu}(U_P - U_P^\dagger)
\vert_{traceless} + \sqrt\epsilon H_\mu(x)
\end{equation}
Here $\epsilon$ is the Langevin time step; 
the sum over $P$ means that $F_\mu$ gets contributions from
all {\sl oriented} plaquettes which include the link $\mu$ at $x$;
the suffix {\sl traceless} means that a subtraction of 
$1/N \Tr{.}$ times the identity matrix
is understood. Finally $H_\mu(x)$ is extracted from a standard
(antihermitian, traceless) Gaussian ensemble of $N$--dimensional
matrices. Now we represent $U_\mu$ by the 
exponential map on $SU(N)$: $U_\mu(x) = \exp\{A_\mu(x)/\sqrt\beta\}$.
The Langevin evolution, when
translated in terms of $A$, is going to depend parametrically on $\beta$.
We then insert a formal power 	series	expansion
\begin{equation}
A_\mu(x) \sim \sum_{k\ge 0} \beta^{-k/2} A_\mu^{(k)}(x)
\end{equation}
into \eq{lange} and we try to match both sides order by
order. To make this possible it is immediately evident that
the time step must be scaled by putting
$\epsilon=\tau/\beta$, which is also
expected since going to large 	$\beta$ requires smaller and smaller
$\epsilon$ to avoid systematic errors. The Langevin algorithm has now been
transformed into an infinite system of coupled non--linear stochastic
finite--difference equations. The system can be 
truncated exactly at the desired order since each field $\Amu{k}$
is independent from higher order fields $A_\nu^{(h)}$ 
with $h>k$; in particular the field $A^{(0)}$
transforms by itself and represents the Abelian limit (a collection of
$(N^2-1)\;$  $U(1)\;$gauge fields). The explicit form of the 
Langevin algorithm at any order can be obtained by expanding
the products of link variables by Baker--Haussdorff--Campbell's
formula.
Since our goal is to reach and, possibly, go beyond
the order $\beta^{-4}$,
we faced the problem of automatically generating these expansions. 
This was achieved using {\sl Mathematica\footnote{
Copyright \copyright\ 1989-94 by Wolfram Research, Inc.}},
which generates the algebraic expansion,
optimizes the code trying to reduce the number of 
operations to a minimum  and finally translates the
output to (parallel) {\tt fortran} or {\tt apese}.
Presently we are running the code designed to reach $O(\beta^{-5})$. 

The explicit form of the expansion 
$F_\mu(x)$ rapidly gets very cumbersome;
to get a rough idea, the number of lines of text in the 
{\sl Mathematica} output starts at 47 at order $\beta^{-2}$ and it 
rises to 3544 at order $\beta^{-4}$.
Hence we report the stochastic equations only for 
the first few orders (already specialized to $SU(3)$)

\begin{eqnarray}
{\Amu{0}}' &=& \Amu{0} - \Fmu{0} \nonumber \\
{\Amu{1}}' &=& \Amu{1} - \Fmu{1}
-\half [ \Fmu{0}, \Amu{0} ] \nonumber \\
{\Amu{2}}' &=& \Amu{2} - \Fmu{2}
-\half [ \Fmu{0}, \Amu{1} ] -\half [ \Fmu{1}, \Amu{0} ] \nonumber \\
&+&{1\over 12} [ \Fmu{0}, [\Fmu{0},\Amu{0}]] 
-{1\over 12} [ \Amu{0}, [\Amu{0},\Fmu{0}]] \nonumber \\
{\Amu{3}}' &=& \Amu{3} - \Fmu{3} 
-\half [ \Fmu{2}, \Amu{0} ] -\half [ \Fmu{1}, \Amu{1} ] 
-\half [ \Fmu{0}, \Amu{2} ] \nonumber \\
&+&{1\over 12} [\Fmu{1},[\Fmu{0},\Amu{0}]] 
+{1\over 12} [\Fmu{0},[\Fmu{1},\Amu{0}]] 
+{1\over 12} [\Fmu{0},[\Fmu{0},\Amu{1}]] \nonumber \\
&-&{1\over 12} [\Amu{0},[\Amu{0},\Fmu{1}]] 
-{1\over 12} [\Amu{0},[\Amu{1},\Fmu{0}]] 
-{1\over 12} [\Amu{1},[\Amu{0},\Fmu{0}]] \nonumber \\
&-&{1\over 24} [\Fmu{0},[\Amu{0},[\Fmu{0},\Amu{0}]]]\,,
\end{eqnarray}
where we denote by $\Fmu{k}(x)$  the 
 expansion  coefficients of the unitary drift 
$F_\mu(x) = \sum_{k\ge 0} \beta^{-(k+1)/2}\;\Fmu{k}(x)$ 
which are explicitly given to the first two orders by

\begin{eqnarray}
\Fmu{0}(x) &=& {\tau\over 6} \sum_{\nu\in P\supset\mu} 
\Anu{\nu}{0}(x) + \sqrt\tau H_\mu(x) \nonumber \\
\Fmu{1}(x) &=& {\tau\over 6}  \sum_{\nu\in P\supset\mu} 
\Anu{\nu}{1}(x) + {\tau\over 12}
\sum_{{\scriptstyle \nu,\rho\in P\supset\mu\atop 
\scriptstyle \nu < \rho}}
[\Anu{\nu}{0}(x),\Anu{\rho}{0}(x)]
\end{eqnarray}
\noindent

Given the stochastic evolution order by order, we consider the expansion
of any observable $\O \sim \sum \beta^{-k/2}\; \O_k$ and we get
\begin{eqnarray*}
\langle\O\rangle &=& \sum\beta^{-k/2} \langle\O_k\rangle \,,\\
\langle\O_k\rangle &=& \lim_{T\to\infty} {1\over T}\sum_{t=1}^T \O_k^{(t)}\,,
\end{eqnarray*}
where $\O_k^{(t)}$ is the value of the random variable,
averaged over the whole lattice, after $t$ steps. 
The existence of the limit in terms of some 
``perturbative probability distribution'' is still to be proven
and it will be  assumed as a working hypothesis here. This fact deserves 
further consideration in view of the fact that at first sight 
the behavior of the stochastic process is far from trivial.

\section{Stochastic Gauge Fixing}\label{GF}
The first run up to level $\beta^{-2}$ was done without any gauge fixing.
The evidence (see \cite{DALLAS}) was rather strong  in favour of
the existence of the time average with an excellent agreement
with the known values; however the statistical fluctuation
of the signal at order 1 and 2 tends to increase in time signaling
the fact that the asymptotic distribution, if it exists at all,
has infinite second moments. The danger of such wild fluctuations
in a numerical calculation is rather obvious  -- as soon as the
signal/fluctuation ratio is smaller than the machine accuracy
the algorithm is dead. While we are not able at present to 
suggest a precise picture of these fluctuations, it is believed
that {\sl fixing the gauge is also going to fix the
problem}. Actually these large fluctuations can be thought of
as gauge--non--invariant contributions to the time trajectory
(with vanishing expectation)
due to the fact that the full gauge invariance is only implemented
by averaging on all possible noise--histories; for a given 
noise history stochastic time evolution and gauge transformations
do not commute and this fact was considered responsible for this
kind of fluctuations in Langevin dynamics (see \cite{PARISI}).
The immediate source of 
divergence can  be identified in  the fields $\Amu{0}$,
which fluctuate like free
Brownian variables in the absence of gauge--fixing. Higher order
fields $\Amu{k}$ contain powers of  $\Amu{0}$ and hence show
stronger fluctuations. Fixing the gauge keeps the fields
$\Amu{0}$ to bounded deviations and hence cures the problem.
The kind of gauge fixing appropriate to Langevin dynamics was
introduced by Zwanziger (\cite{ZWANZIGER}) and implemented as
a lattice algorithm in \cite{ROSSI}. The method consists 
in performing a gauge transformation after each sweep, in such
a way that the field is attracted to the manifold defined by
Landau gauge. The gauge transformation is given by
\begin{eqnarray}
{\cal W}_L: U_\mu(x) &\to& e^{w(x)} U_\mu(x) e^{-w(x+e_\mu)}\nonumber \\
w(x) &=& \alpha\; \sum_\mu\;\Delta_{-\mu}
[U_\mu(x)-U^\dagger_\mu(x)] \Vert_{traceless}\nonumber \\
\Delta_{-\mu}U_\nu(x) &\equiv& U_\nu(x)-U_\nu(x-e_\mu)
\end{eqnarray}
An iteration of ${\cal W}_L$ by itself would bring the 
gauge potential to the Landau gauge. By interleaving it to 
each Langevin step one obtains a sort of soft gauge fixing, 
${\cal W}_L$ providing an additional  drift which however
does not modify the asymptotic probability distribution.
We have now to expand the gauge fixing step ${\cal W}_L$ 
in $1/\sqrt\beta$: this can  be achieved with 
the same technique already developed for the unconstrained
Lengevin algorithm. The parameter $\alpha$ is 
chosen in order to minimize systematic errors --- in practice 
it is set equal to $\tau$.

\section{Results} \label{sec:results}
All the data will refer to the $SU(3)$ simple plaquette 
variable 
\begin{equation}
W = 1-{\scriptstyle {1 \over 3}}\langle\Tr{U_P}\rangle 
\sim \sum \beta^{-n/2} \langle\O_n \rangle = \sum_n c_n \beta^{-n}
\;. 
\end{equation}
The expansion coefficients are obtained from the
time average taken on a trajectory of 2100 Langevin 
steps of which the first 500 were discarded (see Fig.~1).
The odd--degree operators $\O_{2n+1}$, which must have 
vanishing expectation, are used to monitor the process
(see Fig.~2).

The systematic error due to the finite evolution step 
can be corrected by running at different values of $\tau$ and extrapolating
at $\tau=0$ (see Tab.~1).
We check in this way that the method gives very
accurate results when compared to the analytic results, but already
for $c_2$ we need more statistics to get an improvement from
the extrapolation in $\tau$ (see Fig.~3).

\begin{table}[t]
\begin{center}
\begin{tabular}{||c||c|c|c|c||} \hline
$n$ & $1 $ & $2$ & $3$ & $4$   \\ \hline
$c_n (\tau=0.02)$        & 2.019    & 1.210    & 3.001 & 9.58 \\
$\Delta c_n$             & 0.001    & 0.002    & 0.014 & 0.09 \\ 
$c_n (\tau\to 0)$        & 2.000(2) & 1.218(7) & - & -\\
Exact                    & 2        & 1.212(7) & 3.12 & -\\ \hline

\end{tabular}
\caption{Stochastic estimate of perturbative coefficients for the simple 
plaquette, $8^4$ lattice.}
\end{center}
\end{table}

Let us now discuss to what extent our calculation of
the perturbative coefficients is {\sl different} from
those previously published. First of all, of course,
there is an inherent statistical error, due to the
stochastic nature of the calculation, which can be 
improved by accumulating more data.

Other sources of systematic errors, which are
not seen with  the present accuracy, may turn out
to be relevant for other observables. 
First of all there are  finite size effects:
since we have to subtract the perturbative
tail from  Monte Carlo data taken on {\sl the same
finite lattice}, it could turn out that the right thing
to do is {\sl not to correct  for finite size at all}.
For a local quantity like the gluon
condensate the finite size effect is going to be negligible
anyway.  Another difference from the
standard calculation is more fundamental: {\sl we are 
not attempting to extract the leading term in the lattice spacing}.
Our result includes all irrelevant higher order terms in $a$,
which are also present in Monte Carlo data.

\section{Discussion}

Let us now recall from Ref.~\cite{DIGIACOMO} 
that the single plaquette expectation $W_{MC}$ obtained by Monte Carlo 
simulations on the lattice is related to the physical condensate by 
\beq 
W_{MC} = W_{{\rm pert}} + 
{\alpha\over\pi} \langle\bold{F}^2\rangle\; a^4 \; Z(\beta)\,,
\;\;\;\;\;
W_{{\rm pert}} = \sum_{n>0} c_n \beta^{-n}\,,
\eeq
where $W_{{\rm pert}}$ is the lattice perturbative correction 
discussed before. 
To two loop order the condensate in $SU(3)$ is given by the 
asymptotic freedom expression 
\beq\label{as}
G_2^{as}(\beta) \equiv
{\alpha\over\pi} \langle\bold{F}^2 \rangle\; a^4 \; Z(\beta)\,
\simeq
C^{as} \; (\beta/ 6b_0)^{2b_1/b_0^2} e^{ -\beta/3 b_0 }\,
(1-\frac{6 b_1}{b_0\beta}+O(1/\beta^2)) \,
\eeq
where the constant $C^{as} $ is given by the physical quantity
$
C^{as} =\frac{\alpha}{\pi}\langle{\bold F}^2\rangle / \Lambda^4\,,
$
with $\Lambda$ the fundamental lattice gauge theory scale and where
$$
b_0=11/(4\pi)^2\,, \;\;\;\;\;\; b_1=102/(4\pi)^4 \,
$$
are the first two perturbative coefficient of the $\beta$-function.

The main question is then how to extract the 
physical condensate from the Monte Carlo data.
First of all we perform the same analysis done in Ref.~\cite{DIGIACOMO}:
we try to estimate the gluon condensate by comparing 
$G_2^{as}(\beta)$ with the various approximations given by
\beq\label{pert}
G_2^{(n)}(\beta) = W_{MC}-W^{(n)}_{pert} \,,
\;\;\;\;\;\;\;
W^{(n)}_{pert}= \sum^n_{k=1} c_k \beta^{-k} .
\eeq
In Fig.~4 we report $G_2^{(n)}(\beta)$ for $n=1,\ldots,4$ by using 
the values of $c_n$ in Tab. 1  and the Monte Carlo results on the 
single plaquette reported in Ref.~\cite{CDG}.
By increasing the perturbative order $n$ one sees that 
$G_2^{(n)}(\beta)$ has a logarithmic slope in $\beta$ which 
decreases monotonically and seems to approach the 
asymptotic freedom slope in (\ref{as}) given by the dotted  line. 
The series seems still too short for $n=4$. 
Following the same procedure as in Ref.~\cite{DIGIACOMO}, we introduce the 
next two coefficients $c_5$ and $c_6$ which we obtain by fitting 
$G_2^{(6)}(\beta)$ with 
the asymptotic freedom expression $G_2^{as}(\beta)$ in \eq{as}. 
Together with the two coefficients $c_5$, $c_6$ one fits also the 
value $C^{as}$ of the condensate. The resulting points of 
$G_2^{(6)}(\beta)$ are the lower points in Fig.~4 which are 
nicely fitted by the dotted line, the asymptotic freedom expression 
$G_2^{as}(\beta)$.

From the fit one obtains:
$c_5=-84$, $c_6=1193$, and $C^{as}=3.6\times 10^8$. 
This value of $C^{as}$ is close
to the estimate obtained by performing the same analysis at one 
lower loop $n=3$.
Notice that, while the computed coefficients are all 
positive, the first unknown coefficient $c_5$ turns out to be 
negative. This is simply due to the fact that the functions
$\ln G_2^{(n)}(\beta)$ have a slight curvature. 
Indeed, if one applies the same procedure at a lower loop, 
the first fitted coefficient turns out to be negative although 
the known value is positive.

To analyze more closely the possible approach to asymptotic freedom, 
we plot in Fig.~5 the effective slopes of $\ln G_2^{(n)}(\beta)$
computed as follows. We perform a linear fit to $\ln G_2^{(n)}(\beta)$
in various intervals $(\beta_1,\ldots,\beta_2)$ included in 
the interval $6<\beta<7$ where we expect to observe the asymptotic
scaling. The result is shown in Fig.~5, by plotting  
$R=1/slope$ vs. $1/\sqrt n$, the error bars being a measure of the 
different estimate of the slope depending on the interval 
chosen; the asymptotic freedom result 
$R^{as}=3b_0\,$ is plotted as a horizontal line for comparison.
It appears that the values of $R^{(n)}$ have a smooth behaviour 
as function of $1/\sqrt{n}$, which would however extrapolate
to $R^{as}$ for  a finite $n$.
This fact is typical of a divergent series.
Indeed it has been observed \cite{RENORM,ZAK} that this should be the case 
due to the presence of renormalons. The leading renormalon
 gives the following large 
order behaviour of the expansion coefficients
\beq \label{div}
c_n \sim An^B\,n!\; (3 b_0 )^n\,. 
\eeq
As discussed in Ref.~\cite{ZAK}, this behaviour would imply that the 
intrinsic error in the estimate based on the divergent series would 
be of the same order of magnitude
as the asymptotic freedom result $G_2^{as}(\beta)$. 
Actually for $n \le 4$ we find that the coefficients increase 
even faster with $n$ than the one in \eq{div}.
This fact would make even more ambiguous the attempt to extract 
in this manner the gluon condensate from the Monte Carlo data.
Thus one should find some new criterion to identify the perturbative 
subtraction. The stochastic calculation of $c_5$ should be available soon
and we hope it will help in clarifying the situation. 

We conclude by observing that our method can be easily
adapted to evaluate high order perturbative expansions
for any other gauge field observable, in particular 
we may correct for perturbative contributions to Wilson
loops and to the topological susceptibility; the application
to other models, such as chiral models on the lattice,
should also be straightforward.

\newpage\noindent
{\Large\bf Acknowledgments}
\vskip 0.2in
\noindent
 We thank warmly Prof.\ {\bf N.\ Cabibbo, A.\ Di Giacomo, 
A.H.\ Mueller, G.\ Parisi and Pietro Rossi} for useful conversations.
and Prof.\ {\bf G.\ Conte} and the staff of the Computer Center 
of the Faculty of Engineering, University of Parma,
 for granting us computer time on their CM-2.

\begin{figure}[t]
\center{\epsfig{figure=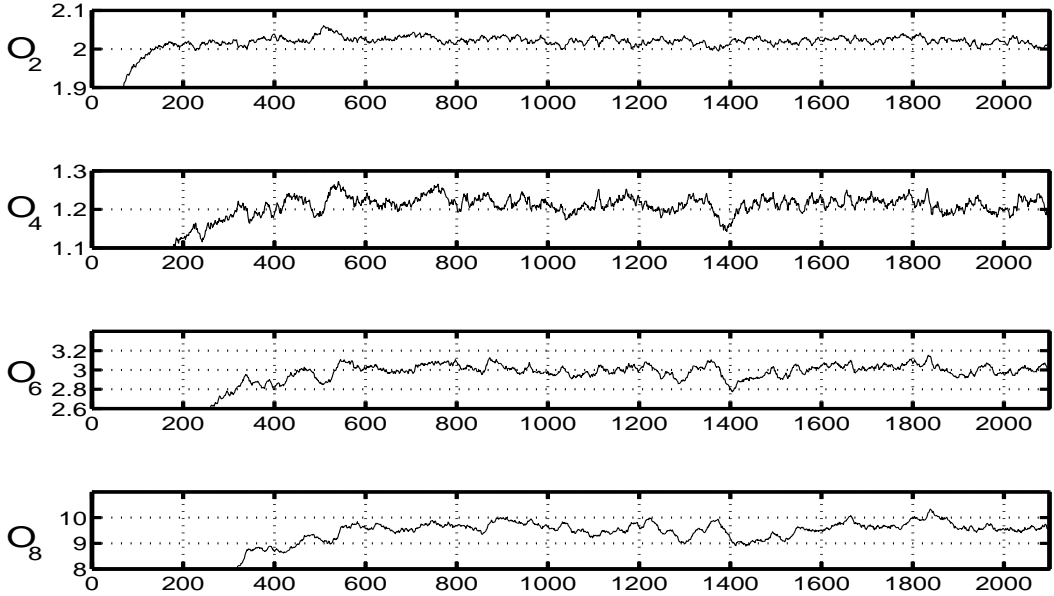,height=8.cm,width=14.cm}}
\caption{Time history of $\O_{2n}, \tau=0.02$ .}
\end{figure}

\begin{figure}[t]
\center{\epsfig{figure=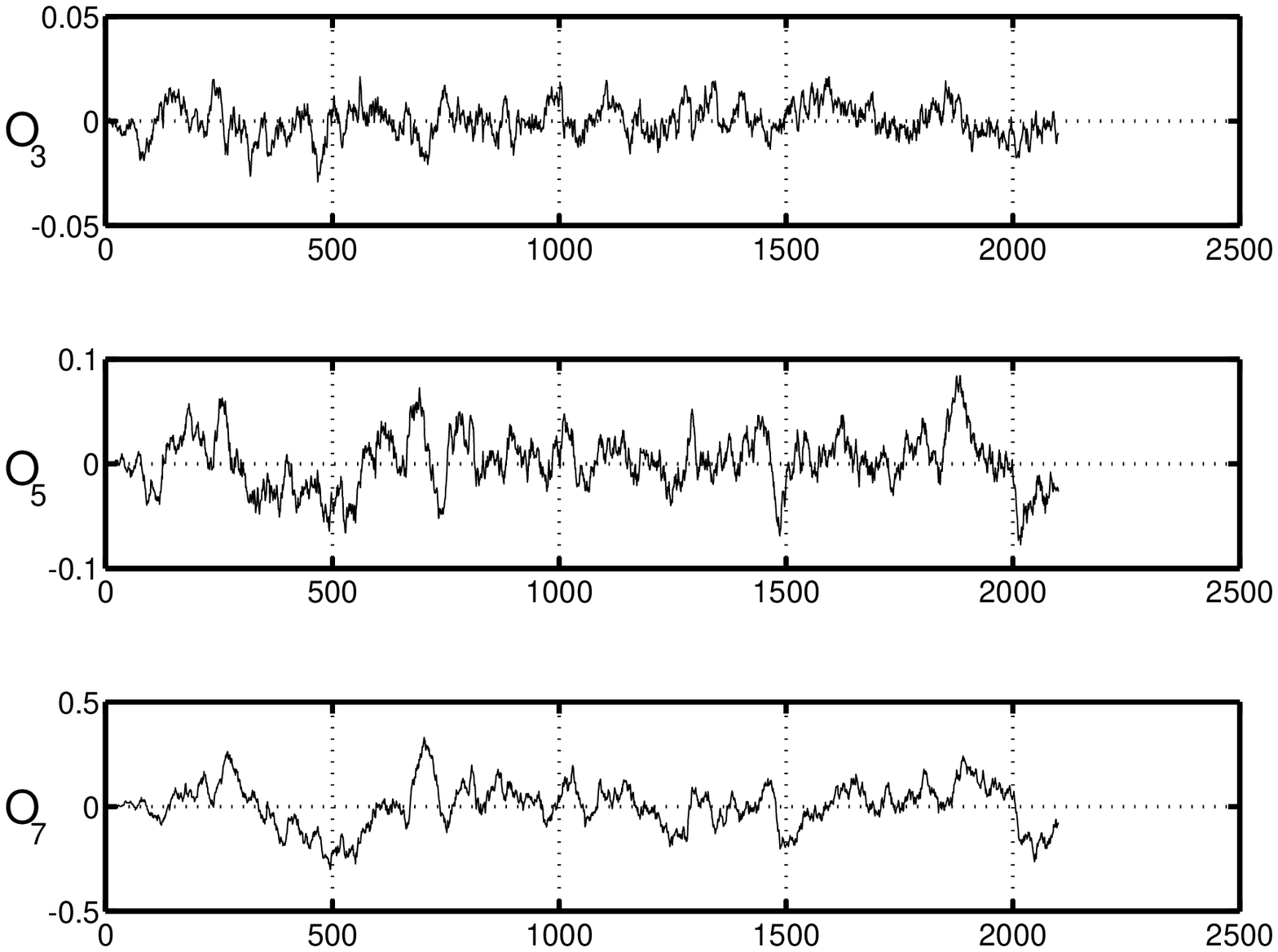,height=6.0cm,width=14.cm}}
\caption{Time history of $\O_{2n+1}. \tau=0.02$ .}
\end{figure}

\begin{figure}[t]\label{extrap}
\center{\epsfig{figure=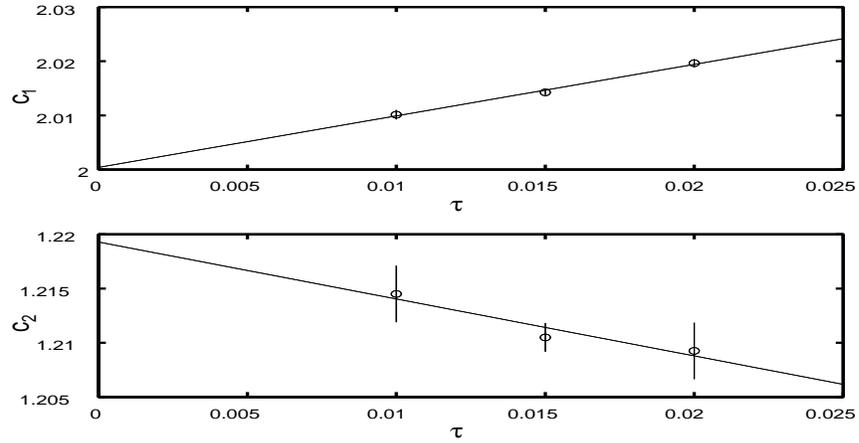,height=6.cm,width=12.cm}}
\caption{ The extrapolation in $\tau$ of the coefficient $c_1, c_2$.}
\end{figure}

\begin{figure}[t]
\center{\epsfig{figure=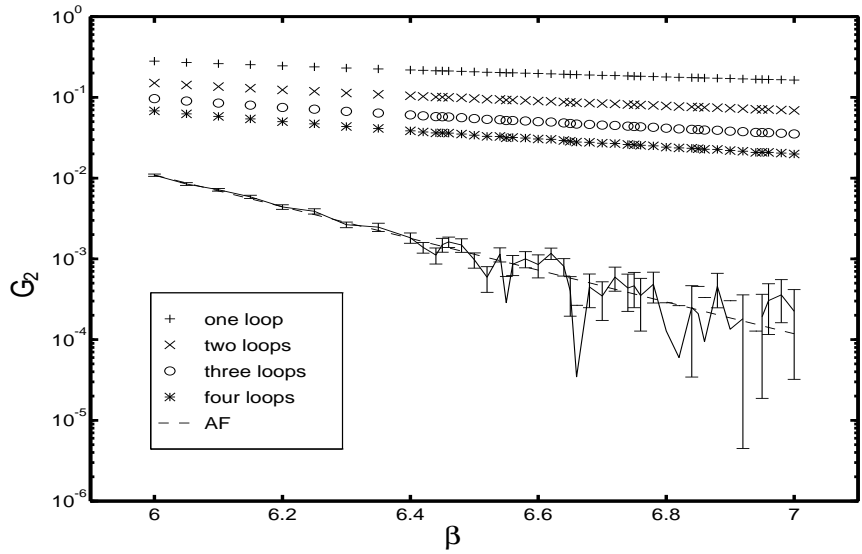,height=8.cm,width=12.cm}}
\caption{ $ G_2^{(n)}$ at various loop orders.
Dashed line is the asymptotic freedom result (\eq{as}) 
with the constant $C^{as}$ obtained by fitting $c_5$ and $c_6$.
The lowest points correspond to $G_2^{(6)}$
with the fitted coefficients $c_5$ and $c_6$.
The error bars refer to the Monte Carlo statistical error only.}
\end{figure}

\begin{figure}[t]\label{slope}
\center{\epsfig{figure=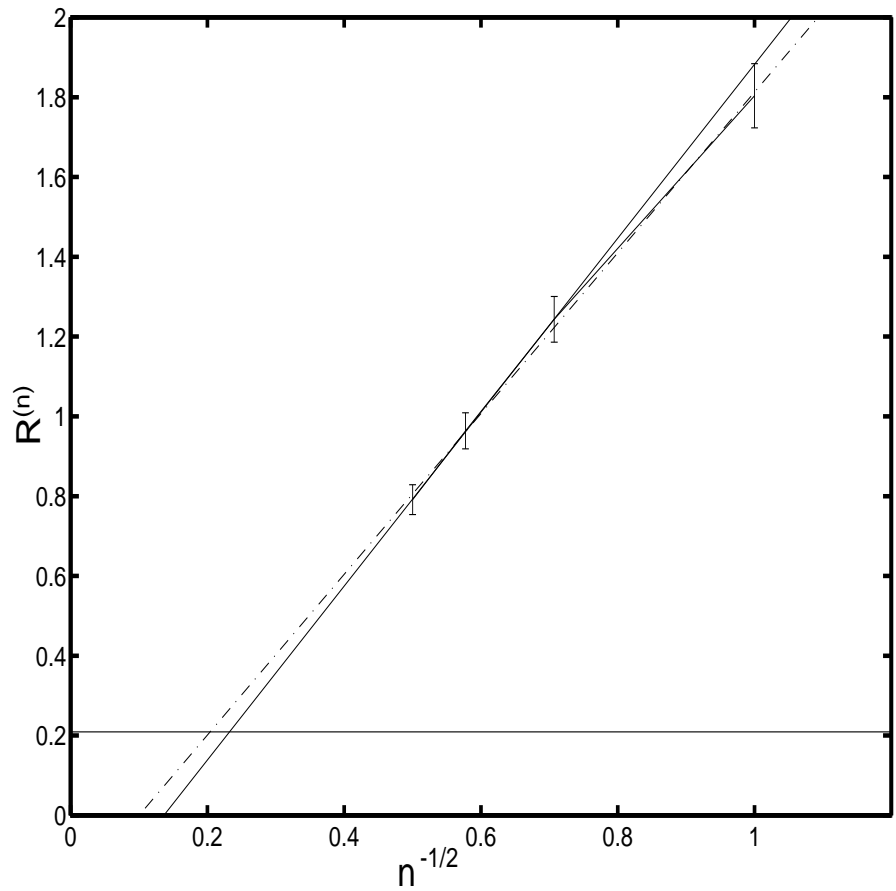,height=12.cm,width=12.cm}}
\caption{ The inverse of the slope of $ \ln(G_2^{(n)})$
vs.  
$1/n^{1/2}$. 
The continuous line 
shows the fit based on the point $n>1$.
The dotted line includes also the point $n=1$. }

\end{figure}

\end{document}